\documentclass[a4paper]{article}
\usepackage{a4,amsmath}

\begin{document}
\Large
\title{\textbf{Relativistic wave equation for one spin-1/2 and one spin-0 particle}}

\author{D.A. Kulikov\thanks{kulikov\_d\_a@yahoo.com},
R.S. Tutik\thanks{tutik@ff.dsu.dp.ua}\\
\\
Theoretical Physics Department \\ Dniepropetrovsk National
University \\
13 Naukova str., Dniepropetrovsk 49050, Ukraine}

\maketitle

\begin{abstract}
\large{ A new approach to the two-body problem based on the
extension of the $SL(2,C)$ group to the $Sp(4,C)$ one is
developed. The wave equation with the Lorentz-scalar and
Lorentz-vector potential interactions for the system of one
spin-$1/2$ and one spin-$0$ particle with unequal masses is
constructed.}
\end{abstract}


\section{Introduction}
\label{part1} The relativistic two-body problem has numerous
applications in particle and nuclear physics. Because the
Bethe-Salpeter equation \cite{bethe} is exceedingly difficult to
solve, even numerically, different approaches to this problem have
been developed. They include: reductions of the Bethe-Salpeter equation resulting in the
quasipotential approach \cite{logunov} and the Breit-type
equations \cite{breit,bijtebier,tanaka};
relativistic quantum mechanics with constraints
\cite{sazdjian86,crater87,crater06} that uses a system of two
coupled equations describing individual particles;
the Barut method \cite{barut,klimek} for deriving a
single two-body wave equation from a field-theoretical action;
Lorentz-invariant two-body wave equations having the
Schr\"odinger-like \cite{fushchych} or Dirac-like \cite{nikitin}
form.

In the last works the wave functions transform according to the
more complicated representations than the one-particle wave
functions that can be regarded as involving the extended Lorentz
symmetry. The explicit extensions of the Lorentz group, including
the symplectic \cite{pirogov} and the general complex
\cite{bogush} ones, have been studied, too.

Recently, the extension of the $SL(2,C)$ group to the $Sp(4,C)$ one
has allowed us to formulate a new approach to the relativistic two-body problem \cite{kulikov}. The goal of the present work is
to apply this technique for constructing the wave equation
for the system composed of the spin-$1/2$ and spin-$0$ particles with unequal masses.

\section{Symplectic space-time extension}
\label{part2} The relativistic theory is usually formulated in the
Minkowski space with the homogeneous Lorentz group $SO(1,3)$ as
the local symmetry group. However,  since the $SO(1,3)$ group is
covered by the $SL(2,C)\equiv Sp(2,C)$ group, the relativistic
field theory can equivalently be formulated entirely within the
framework of the $Sp(2,C)$ Weyl spinors \cite{penrose}.

Recall that the symplectic $Sp(2l,C)$ group is the
group of $2l\times 2l$ matrices with complex elements and
determinant equal to one \cite{weyl}. These matrices act on
$2l$-component Weyl spinors and preserve an antisymmetric bilinear
form which plays the role of "metrics" in the spinor space. For the
$Sp(4,C)$ group, we denote this form by
$\eta_{\alpha\beta}=-\eta_{\beta\alpha}$ ($\alpha,\beta=1,2,3,4$).
Then the $Sp(4,C)$ Weyl spinors
$\varphi_{\alpha}$ with lower indices and their complex
conjugatives $\bar{\varphi}_{\bar{\alpha}}=(\varphi_{\alpha})^*$
are related to spinors with upper indices by transformations
$\varphi_{\alpha}=\eta_{\alpha\beta}\varphi^{\beta}$ and
$\bar{\varphi}_{\bar{\alpha}}=\eta^*_{\bar{\alpha}\bar{\beta}}\bar{\varphi}^{\bar{\beta}}$.

Further, there exists one-to-one correspondence between $Sp(2l,C)$
Hermitian spin-tensors of second rank and $(2l)^2$-component real
vectors. In the case of the $Sp(2,C)$ group, they are ordinary
Minkowski four-vectors. For the case of $Sp(4,C)$ group, we define
the correspondence between the Hermitian spin-tensor,
$P_{\alpha\bar{\alpha}}$, and a real vector $P_M$ by
\begin{equation}\label{eq1}
P_{\alpha\bar{\alpha}}=\mu^M_{\alpha\bar{\alpha}}P_M, \qquad
P^M=\frac{1}{4}\tilde{\mu}^{M\bar{\alpha}\alpha}P_{\alpha\bar{\alpha}}
\end{equation}
where $\mu^M_{\alpha\bar{\alpha}}$ ($M=1\div 16$) are matrices of
the basis in the space of $4 \times 4$ Hermitian matrices and
tilde labels the transposed matrix with uppered spinor indices.
In what follows, the spinor indices will be omitted when
possible.

To clarify the relationship between the discussed vector space and the
Minkowski space $\mathsf{R^{4}}$, we represent 16 values of the
vector index of $P_M$ through $4\times 4$ combinations of two
indices, $M=(a,m)$, with both $a$ and $m$ running from $0$ to $3$.
Then the metrics of the discussed vector space
is reduced to the factorized form
\begin{equation}\label{eq3}
g^{MN}\equiv g^{(a,m)(b,n)}=\hat{h}^{ab}h^{mn}
\end{equation}
where $h^{mn}=diag(1,-1,-1,-1)$ is the usual Minkowski metrics and
$\hat{h}^{ab}=diag(1,1,-1,1)$ is caused by the group extension.

The factorization of the metrics means that the vector from
$\mathsf{R^{16}}$ may be decomposed into four Minkowski
four-vectors. As a consequence, we can use these 16-component
vectors or, equivalently,  $Sp(4,C)$ Hermitian spin-tensors to
construct the wave equation for a few-body system.

\section{Wave equation for a fermion-boson system}
\label{part3}

Let us consider a system consisted of one
spin-1/2 and one spin-0 particle. With the total
spin of the system being equal to $1/2$, the wave equation must
have the form of the Dirac-like equation in which the wave function
is represented by a Dirac spinor or,
in our case, by two $Sp(4,C)$ Weyl spinors as
\begin{equation}\label{eq4}
P_{\alpha\bar{\alpha}}\bar{\chi}^{\bar{\alpha}}=m\varphi_{\alpha},
\qquad
\tilde{P}^{\bar{\alpha}\alpha}\varphi_{\alpha}=m\bar{\chi}^{\bar{\alpha}}
\end{equation}
where $P_{\alpha\bar{\alpha}}$ is the $Sp(4,C)$ momentum
spin-tensor and $m$ is a mass parameter. According to the splitting of the vector indices,
we have
\begin{equation}\label{eq5}
P=\mu^{(a,m)}P_{(a,m)}=\Sigma^0\otimes\sigma^m
w_m+\Sigma^1\otimes\sigma^m p_m+ \Sigma^2\otimes\sigma^m
r_m+\Sigma^3\otimes\sigma^m q_m
\end{equation}
where $w_m$, $p_m$, $r_m$, $q_m$ are the Minkowski four-momenta
and matrices $\Sigma^a$, $\sigma^m$ may be expressed in terms of
$2\times 2$ unit matrix $I$ and the Pauli matrices $\tau^i$.

It has been shown \cite{kulikov} that the wave equation
(\ref{eq4}) describes the fermion-boson system with the equal mass
constituents. Now we are going to generalize it to the case of the
particles with unequal masses. For this purpose, let us replace
the mass parameter in the right hand of Eq.(\ref{eq4}) by a
suitable matrix term which can be expressed as a combination of
direct products of matrices. Though such term breaks the $Sp(4,C)$
symmetry of the wave equation, but the Lorentz $SO(1,3) \subset
Sp(4,C)$ symmetry is retained. It becomes obvious if the second
matrix in the direct product is chosen as a unit matrix and the
first one is written through the matrices $\Sigma^a$, like in
Eq.(\ref{eq5}). There are two equivalent possibilities, with the
matrix $\Sigma^a$ chosen as $\Sigma^1=\tau^1$ or $\Sigma^3=\tau^3$
($\Sigma^0 =I$ is the trivial choice), that result in the plus
sign in the metrics $\hat{h}^{ab}$ defined by Eq.(\ref{eq3}). In
view of this we replace the mass parameter as follows
\begin{equation}\label{eq100}
m\rightarrow(m_1+m_2)/2+\tau^1\otimes I (m_1-m_2)/2,
\end{equation}
so that the additional term vanishes if $m_1=m_2$.

Thus, the wave equation for the fermion-boson system with unequal masses takes the form
\begin{equation}\label{eq11}
P\bar{\chi}=(m_{+}+\tau^1\otimes I m_{-})\varphi, \qquad
\tilde{P}\varphi=(m_{+}+\tau^1\otimes I m_{-})\bar{\chi}
\end{equation}
where $m_{\pm}=(m_{1}\pm m_{2})/2$.

Now, for elucidating the two-particle interpretation of the proposed equation,
we consider the structure of the the $Sp(4,C)$ momentum spin-tensor
given by Eq.(\ref{eq5}). It should be stressed that the description of the two-particle system requires only
two four-momenta whereas the $Sp(4,C)$ momentum spin-tensor
corresponds to four four-momenta, collected in a $16$-component
vector. Therefore the number of the
independent components of $w_m$, $p_m$, $r_m$ and $q_m$ must be decreased
that can be implemented with subsiduary conditions.

In order to derive the subsiduary conditions we transform Eq.(\ref{eq11}) to the form of the
Klein-Gordon equation. By eliminating $\bar{\chi}$ and using Eq.(\ref{eq5}), we obtain
\begin{equation}\label{eq6}
(w^2+p^2-r^2+q^2-\frac{2m_{-}}{m_{+}}wp-m_{+}^2+m_{-}^2+\sum^{5}_{A=1}\gamma_A
K^A)\varphi=0
\end{equation}
where $w^2=(w^0)^2-\mathbf{w}^2$, $p^2=(p^0)^2-\mathbf{p}^2$ etc,
$\gamma_A$ are direct products of the Pauli matrices, and $K^A$
are quadratic forms with respect to the four-momenta.

Because in this equation the non-diagonal terms $\gamma_A K^A$
have no analog in the case of the ordinary Klein-Gordon equation,
we put $\gamma_A K^A=0$ that yields
\begin{eqnarray}\label{eq12}
&&(m_{+}^{2}-m_{-}^{2})(wp-m_{+}m_{-})-m_{+}m_{-}(r^2-q^2)=0, \nonumber \\
&&m_{+}wq-m_{-}pq=0, \nonumber \\
&&m_{+}rp-m_{-}rw=0,  \\
&&rq=0, \nonumber \\
&&m_{+}(r^m w^n - r^n w^m -\epsilon^{mnkl}p_k q_l)-m_{-}(r^m p^n -
r^n p^m -\epsilon^{mnkl}w_k q_l)=0, \nonumber
\end{eqnarray}
with $\epsilon^{mnkl}$ being the totally antisymmetric tensor
($\epsilon^{0123}=+1$).

Thus, the imposed conditions and the
Klein-Gordon-like equation set ten components of $w_m$, $p_m$,
$r_m$, $q_m$ to be the independent ones. For the connection of these
four-momenta with the four-momenta, $p_{1m}$ and
$p_{2m}$, of the constituent particles we assume
\begin{equation}\label{eq8}
w_m=\frac{1}{2}(p_{1m}+p_{2m}),\quad
p_m=\frac{1}{2}(p_{1m}-p_{2m}),\quad r_m=0,\quad q_m=0.
\end{equation}
Then the only one condition from Eqs.(\ref{eq12}) remains non-trivial that reads
\begin{equation}\label{eq13}
(wp-m_{+}m_{-})\equiv (p_1^2-p_2^2-m_1^2+m_2^2)/4=0.
\end{equation}
This equality implies that the total spinor wave function does not
depend on the relative time of the particles.

Further, the wave equation (\ref{eq11}) and the condition
(\ref{eq13}) can be reduced to the one-particle Dirac and
Klein-Gordon equations for the constituents of our system. Indeed,
with decomposing the spinor wave functions into the projections
\begin{equation}\label{eq14}
\varphi_{\pm}=\frac{1}{2}(1\pm\tau^1\otimes I)\varphi, \qquad
\bar{\chi}_{\pm}=\frac{1}{2}(1\pm\tau^1\otimes I)\bar{\chi}
\end{equation}
which are two-component
$Sp(2,C)$ Weyl spinors as well, Eqs.(\ref{eq11}) and (\ref{eq13}) reduce to
two uncoupled sets of equations
\begin{equation}\label{eq15}
p_{1m}\sigma^m\bar{\chi}_{+}=m_1\varphi_{+}, \qquad
p_{1m}\tilde{\sigma}^m\varphi_{+}=m_1\bar{\chi}_{+}
\end{equation}
\begin{equation}\label{eq15a}
(p_2^2-m_2^2)\varphi_{+}=0, \qquad (p_2^2-m_2^2)\bar{\chi}_{+}=0
\end{equation}
and
\begin{equation}\label{eq16}
p_{2m}\sigma^m\bar{\chi}_{-}=m_2\varphi_{-}, \qquad
p_{2m}\tilde{\sigma}^m\varphi_{-}=m_2\bar{\chi}_{-},
\end{equation}\begin{equation}\label{eq16a}
(p_1^2-m_1^2)\varphi_{-}=0, \qquad (p_1^2-m_1^2)\bar{\chi}_{-}=0,
\end{equation}
consisted of the free one-particle Dirac equations written in the Weyl
spinor formalism \cite{landau} and the free Klein-Gordon equations.

Hence it appears
that the wave equation (\ref{eq11}) supplemented with the
subsiduary conditions (\ref{eq12}) describes two systems composed
of the spin-$1/2$ and spin-$0$ particles. These systems differ
from each other only in permutation of masses of the particles.
As a next step, we must include the potential interaction in our
equations.

\section{Inclusion of potential interaction}
\label{part4}

A generally accepted receipt of introducing the interaction
consists in the replacement of the four-momenta of the particles in the
minimal manner by the generalized momenta
($p_i^m\rightarrow\pi_i^m=p_i^m-A_i^m$, i=1,2), so that each
particle is in an external potential of the other. This kind of
coupling is referred to as the Lorentz-vector interaction. Another
possibility uses the mass-potential substitution,
$m_i\rightarrow m_i+S_i$, that corresponds to the Lorentz-scalar
interaction.

In our approach the masses and four-momenta of the particles are
involved through the quantities $w^m, p^m, m_{+}, m_{-}$. For this
reason, we introduce the Lorentz-vector and Lorentz-scalar interactions
by the replacements
\begin{eqnarray}\label{eq17}
&&w^m\rightarrow\omega^m=w^m-A^m, \qquad
m_{+}\rightarrow M_{+}=m_{+}+S_{+}, \nonumber \\
&&p^m\rightarrow\pi^m=p^m-B^m, \qquad m_{-}\rightarrow
M_{-}=m_{-}+S_{-}.
\end{eqnarray}

Here the involved potentials $A^m, B^m, S_{+}, S_{-}$ may
depend on the coordinates and momenta of the particles
but the shape of these potentials is restricted. This restriction is
caused by the requirement that the wave equation must be compatible with
the subsiduary condition (\ref{eq13}) written after the
replacements (\ref{eq17}) as
\begin{equation}\label{eq183}
L\equiv \omega\pi+\pi\omega-M_{+}M_{-}-M_{-}M_{+}=0
\end{equation}

A sufficient condition for this compatibility is that the operator $L$ of the subsiduary
condition should commute with the operators in the wave equation:
\begin{equation}\label{eq184}
[L,\omega_m]\approx 0, \quad [L,\pi_m]\approx 0,
\quad[L,M_{+}]\approx 0, \quad[L,M_{-}]\approx 0
\end{equation}
where the weak equality sign means that the commutator may give an
expression proportional to $L$ itself which, on account of Eq.(\ref{eq183}),
equals to zero.

Because for the quantity $wp$ appearing in Eqs.(\ref{eq17}) and
(\ref{eq183}), we have $[wp,x_m]\neq 0$ but $[wp,x_{\bot m}]=0$,
the conditions in Eqs.(\ref{eq184}) require that the potentials
depend on the relative coordinate $x_m=x_{1m}-x_{2m}$ only through
its transverse with respect to the total momentum part
\begin{equation}\label{eq185}
x_{\bot}^m=(h^{mn}-w^m w^n/w^2)x_n
\end{equation}
where the total momentum $w_m$ is assumed to be a constant of
motion.

The simplest solution to the compatibility condition (\ref{eq184}) comes from
the following ansatz
\begin{equation}\label{eq186}
\omega\pi+\pi\omega=2Cwp, \qquad
M_{+}M_{-}+M_{-}M_{+}=2Cm_{+}m_{-},
\end{equation}
where $C=C(x_{\bot}^2)$ is an arbitrary function. Then the
subsiduary condition (\ref{eq183}) takes the form of
Eq.(\ref{eq13}), which describes the case without interaction,
that brings at once to vanishing commutators.

Finally, let us derive an explicit form of the wave equation
for the fermion-boson system with the potential interactions.
With substituting the generalized momenta and the mass-potential
terms (\ref{eq17}) into Eqs.
(\ref{eq5}) and (\ref{eq11}), we obtain
\begin{eqnarray}\label{eq182}
&&(I\otimes \sigma^m \omega_m+\tau^1\otimes \sigma^m
\pi_m)\bar{\chi}
=(M_{+}+\tau^1\otimes I M_{-})\varphi, \nonumber \\
&&(I\otimes \tilde{\sigma}^m \omega_m+\tau^1\otimes
\tilde{\sigma}^m \pi_m)\varphi=(M_{+}+\tau^1\otimes I
M_{-})\bar{\chi}.
\end{eqnarray}

Here the quantities $\omega_m, \pi_m, M_{+}, M_{-}$ involve
the interaction and satisfy the ansatz (\ref{eq186}).
Using this ansatz, we can introduce both the potential interaction
described by the time-component of the Lorentz vector and
the confinement potential included in the Lorentz-scalar term or
in the spatial part of the Lorentz vector.

Thus, a new approach to the two-body problem based on the
extension of the $SL(2,C)$ group to the $Sp(4,C)$ one has been
developed. It permits us to construct the relativistic
wave equation for the system consisted of spin-$1/2$ and spin-$0$
particles with unequal masses, involving the various forms of interaction.

\end{document}